\documentclass{PoS}
\usepackage{amsmath}
\usepackage{amssymb}
\usepackage{graphicx}
\usepackage{cite}
\usepackage{tabularx}
\title{Full QCD in external chromomagnetic field}

\ShortTitle{Full QCD in external chromomagnetic field}

\author{Paolo Cea \\ Universit\`a di Bari  \& INFN - Bari, \\
E-mail:  \email{paolo.cea@ba.infn.it}}

\author{\speaker{Leonardo Cosmai} \\ INFN - Bari \\
E-mail:  \email{leonardo.cosmai@ba.infn.it}}

\author{Massimo D'Elia  \\  Universit\`a di Genova \& INFN - Genova \\
E-mail:  \email{massimo.delia@ge.infn.it} }

\abstract{ We investigate the deconfining phase transition in full
QCD with two flavors of staggered fermions in presence of a
constant abelian chromomagnetic field. We find that the
deconfinement temperature decreases and eventually goes to zero by
increasing the strength of the chromomagnetic field. Moreover our
results suggest that the chiral transition coincides with the
deconfinement transition and therefore even the chiral critical
temperature depends on the applied chromomagnetic field. We also
find that the chiral condensate increases with the strength of the
chromomagnetic field.}

\FullConference{XXIV International Symposium on Lattice Field Theory\\
         July 23-28 2006\\
         Tucson Arizona, US}

\begin{document}

\newcommand{\be}{\begin{equation}}
\newcommand{\ee}{\end{equation}}

%%%%%%%%%%%%%%%%%%%%%%%%%%%%%%%%%%%%%%%%%%%%%%%%%%%%%%%%%%%%%%%%%%%%%%%
\section{Introduction}
\label{Introduction}
In a previous study~\cite{Cea:2005td}  on the vacuum dynamics of non abelian gauge theories we found that
the deconfinement temperature depends on the strength of an external abelian chromomagnetic field.
In particular we found that
the deconfinement temperature decreases when the strength of the applied field is increased and eventually goes to zero.
This effect corresponds to the reversible Meissner effect in the case of ordinary superconductors therefore
we refer to it as "vacuum color Meissner effect".
This effect could shed light on confinement/deconfinement dynamics, therefore in our opinion it is important
to test if this effect continues to hold even when switching on fermionic degrees of freedom.

The aim of the present work  is to understand if
the deconfinement temperature depends
on the strength of an external abelian chromomagnetic field
even  in the case of  full QCD.

To this purpose  we performed numerical simulations for
finite temperature $N_f=2$  QCD in an external abelian chromomagnetic field.
Simulations have been done using APEmille crate in Bari and
the recently installed computer facilities at
INFN apeNEXT Computing Center in Rome.

\section{The method}
\label{Themethod}
To investigate the QCD dynamics in presence of a external background field at finite temperature
we use the following lattice thermal partition functional~\cite{Cea:1997ff,Cea:1999gn,Cea:2004ux}
\begin{eqnarray}
\label{ZetaT}
\mathcal{Z}_T \left[ \vec{A}^{\text{ext}} \right]  &=
&\int_{U_k(L_t,\vec{x})=U_k(0,\vec{x})=U^{\text{ext}}_k(\vec{x})}
\mathcal{D}U \,  {\mathcal{D}} \psi  \, {\mathcal{D}} \bar{\psi} e^{-(S_W+S_F)}
\nonumber \\&=&  \int_{U_k(L_t,\vec{x})=U_k(0,\vec{x})=U^{\text{ext}}_k(\vec{x})}
\mathcal{D}U e^{-S_W} \, \det M \,,
\end{eqnarray}
where $S_W$ is the Wilson action, $S_F$ is the fermion action, and $M$ is the fermionic matrix.
The spatial links  are constrained to values corresponding to the external background field, whereas
the fermionic fields are not constrained.
The relevant quantity is the free energy functional  defined as
\begin{equation}
{\mathcal{F}}[\vec{A}^{\text{ext}}] = -\frac{1}{L_t} \ln
\left\{
\frac{
{\mathcal{Z}}_T[\vec{A}^{\text{ext}}]}
{{\mathcal{Z}}_T[0]}
\right\}   \,.
\end{equation}
We can evaluate by numerical simulations the derivative of the free energy functional with respect to the gauge coupling
\begin{equation}
\label{deriv}
F^\prime(\beta) =
\frac{\partial {\mathcal{F}}(\beta)}{\partial \beta} =
V  \left[
<U_{\mu\nu}>_{\vec{A}^{\text{ext}}=0} -
<U_{\mu\nu}>_{\vec{A}^{\text{ext}} \ne 0}
\right] \equiv V f(\beta) \,,
\end{equation}
where the subscripts on the averages indicate the value of the external field.
The generic plaquette
$U_{\mu\nu}(x)=U_\mu(x)U_\nu(x+\hat{\mu})U^\dagger_\mu(x+\hat{\nu})U^\dagger_\nu(x)$
contributes to the sum in Eq.~(\ref{deriv}) if the link $U_\mu(x)$
is a "dynamical" one, i.e. it is not constrained in the functional
integration eq.~(\ref{ZetaT}).
Observing that $f[\vec{A}^{\text{ext}}] = 0$ at $ \beta = 0$, we
may eventually obtain  $f[\vec{A}^{\text{ext}}]$ from
$f^{\prime}[\vec{A}^{\text{ext}}]$ by numerical integration:
\be
\label{trapezu1}
f[\vec{A}^{\text{ext}}]  =  \int_0^\beta
f^{\prime}[\vec{A}^{\text{ext}}] \,d\beta^{\prime} \; .
\ee

\section{Numerical results}
\label{numericalresults}
We consider an external abelian chromomagnetic field directed along spatial direction 3 and color direction 3.
On the lattice this corresponds to constrain the spatial links to
\begin{equation}
\begin{split}
& U^{\text{ext}}_1(\vec{x}) =
U^{\text{ext}}_3(\vec{x}) = {\mathbf{1}} \,,
\\
& U^{\text{ext}}_2(\vec{x}) =
\begin{bmatrix}
\exp(i \frac {g H x_1} {2})  & 0 & 0 \\ 0 &  \exp(- i \frac {g H
x_1} {2}) & 0
\\ 0 & 0 & 1
\end{bmatrix}
\end{split}
\end{equation}
where due to toroidal topology of the lattice the field strength gets
quantized
\begin{equation}
\label{a2gH}
a^2 \frac{g H}{2} = \frac{2 \pi}{L_1}
n_{\text{ext}} \,\,\,, n_{\text{ext}}\,\,\,{\text{integer}} \,.
\end{equation}
In Fig.~\ref{Fig1} we display the derivative of the free energy and the chiral condensate at
fixed external field strength.
%
%
%
% FIGURE 1
\FIGURE[ht]{\label{Fig1}
\includegraphics[width=0.75\textwidth,clip]{fig_01.eps}
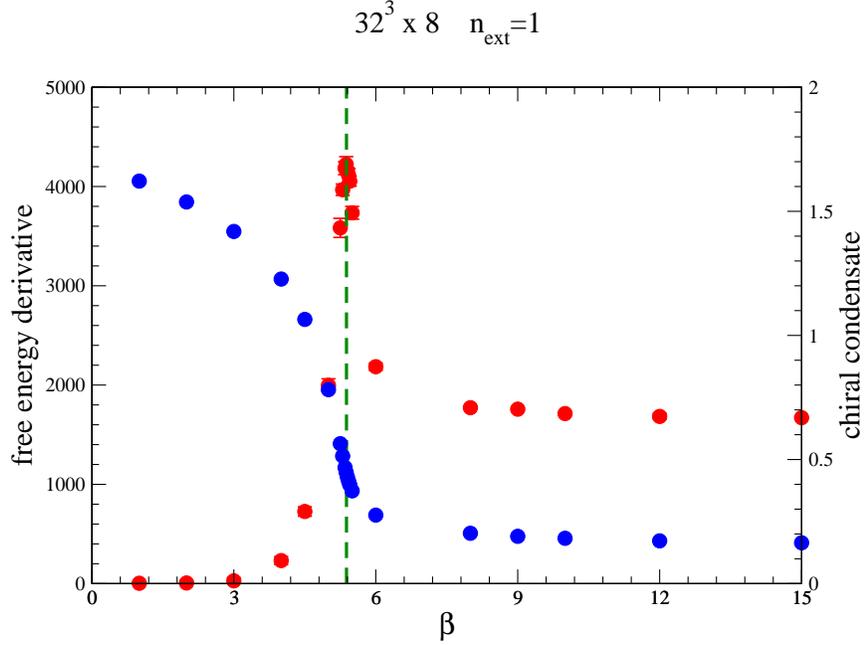
\caption{The derivative of the free energy (blue circles) and the chiral condensate (red circles)
vs. $\beta$ on a $32^3\times8$ lattice and bare quark mass $am_q=0.075$
in correspondence of external field strength $n_{\text{ext}}=1$.}}
The peak in the derivative of the free energy correlates to the drop of the chiral condensate.

The deconfinement phase transition is located by looking at the peak of the derivative of the free energy.
In pure gauge theories we found~\cite{Cea:2005td} that the peak position depends on the strength of the external
chromomagnetic field. Indeed by varying the external field strength we found even in the case of full QCD
that the phase transition shifts by varying the field strength. In Fig.~\ref{Fig2} the free energy derivative vs. $\beta$
in correspondence of three values of the field strength is displayed. And one can see that the peak of the free energy derivative
shifts towards lower values of $\beta$ by increasing the field strength.
%
%
%
% FIGURE 2
\FIGURE[ht]{\label{Fig2}
\includegraphics[width=0.75\textwidth,clip]{fig_02.eps}
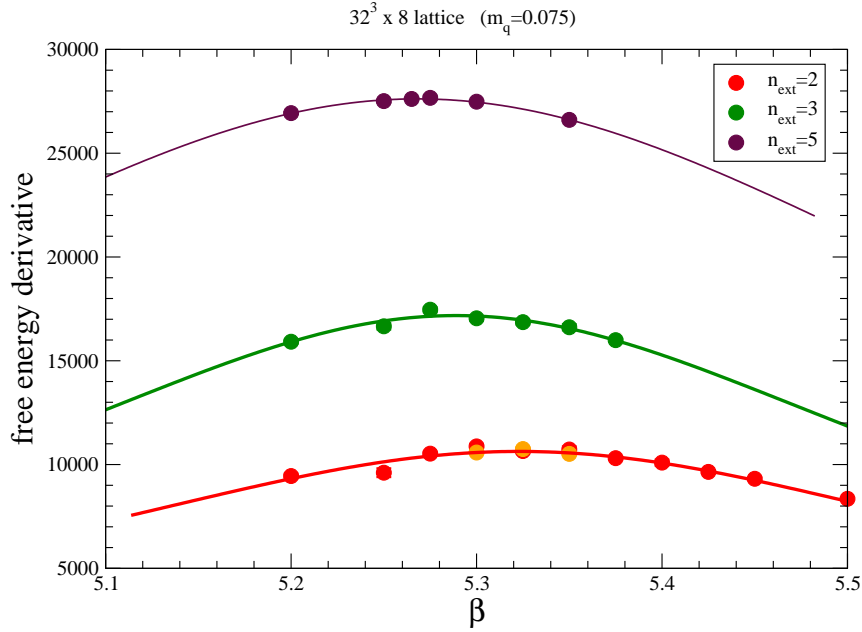
\caption{The peak position in the derivative of the free energy
in correspondence of three values of the external field strength.}}
It is also interesting to notice that (see Fig.~\ref{Fig3}) the value of the chiral condensate
depends on the strength of the applied chromomagnetic field.
%
%
%
% FIGURE 3
\FIGURE[ht]{\label{Fig3}
\includegraphics[width=0.75\textwidth,clip]{fig_03.eps}
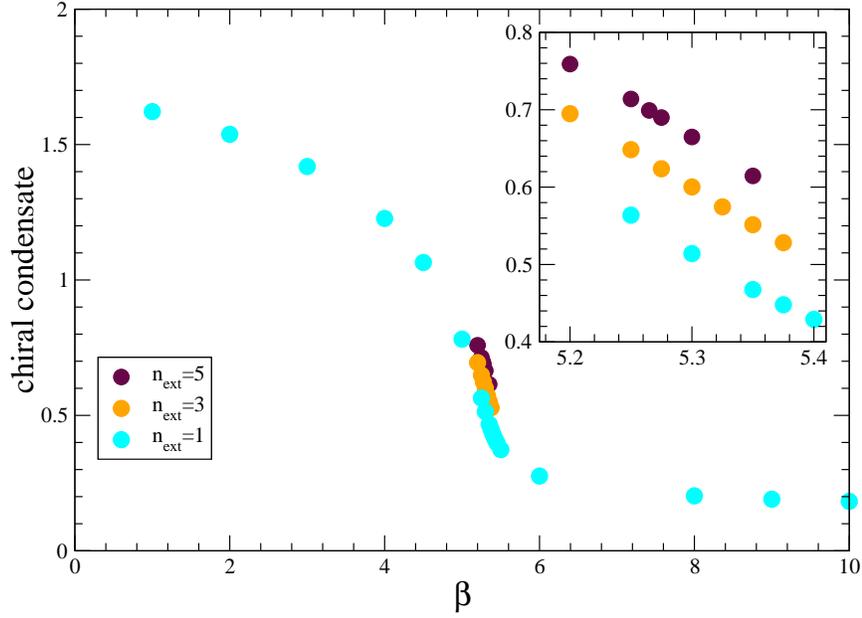
\caption{The chiral condensate vs. $\beta$ in correspondence of different values of the external field
strength. The region near the chiral phase transition is displayed in the inset .}}
%
%
%
%
%
% FIGURE 4a
\FIGURE[ht]{\label{Fig4a}
\includegraphics[width=0.75\textwidth,clip]{fig_04a.eps}
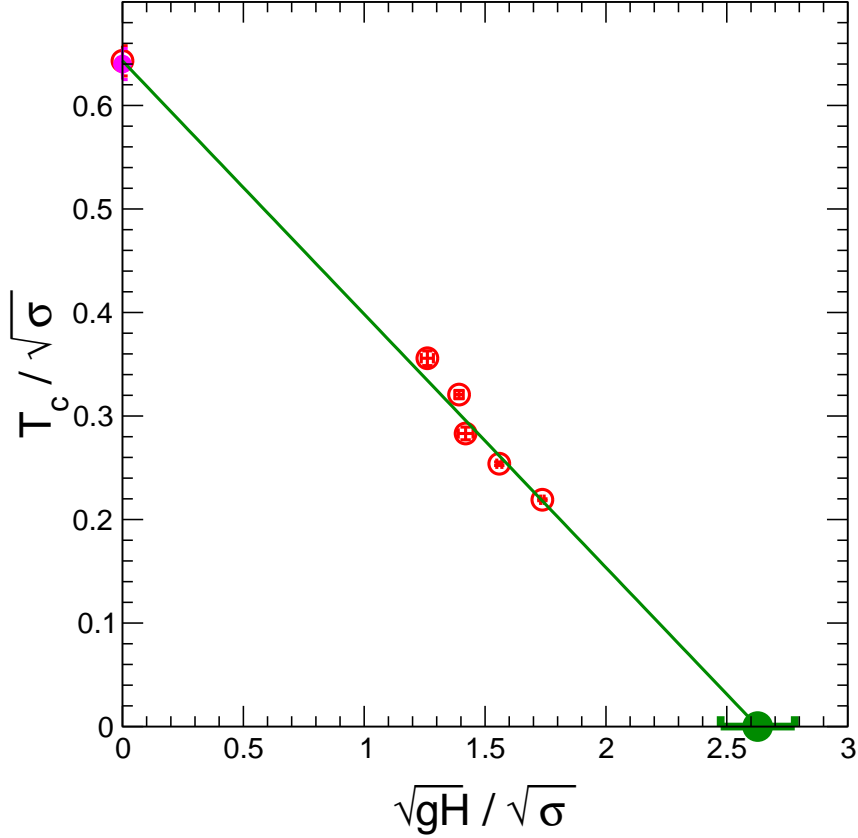
\caption{SU(3) pure gauge. The critical temperature $T_c$ estimated on a $64^3 \times 8$ lattice in units of
the string tension, eq.~(\ref{Tphys}), versus the square root of the field strength
$\sqrt{gH}$ in units of the string tension, eq.~(\ref{gHphys}). Solid line is the linear fit eq.~(\ref{tcsqrtsigma}).
In correspondence of zero vertical axis: open circle  is $T_c/\sqrt{\sigma}$ at zero external field;
full circle is the determination of $T_c/\sqrt{\sigma}$ obtained in the literature~\cite{Teper:1998kw}
The green full point is the critical field in units of the string tension, eq.~(\ref{su3hc}).}}
%
%
%
%
%
% FIGURE 4b
\FIGURE[ht]{\label{Fig4b}
\includegraphics[width=0.75\textwidth,clip]{fig_04b.eps}
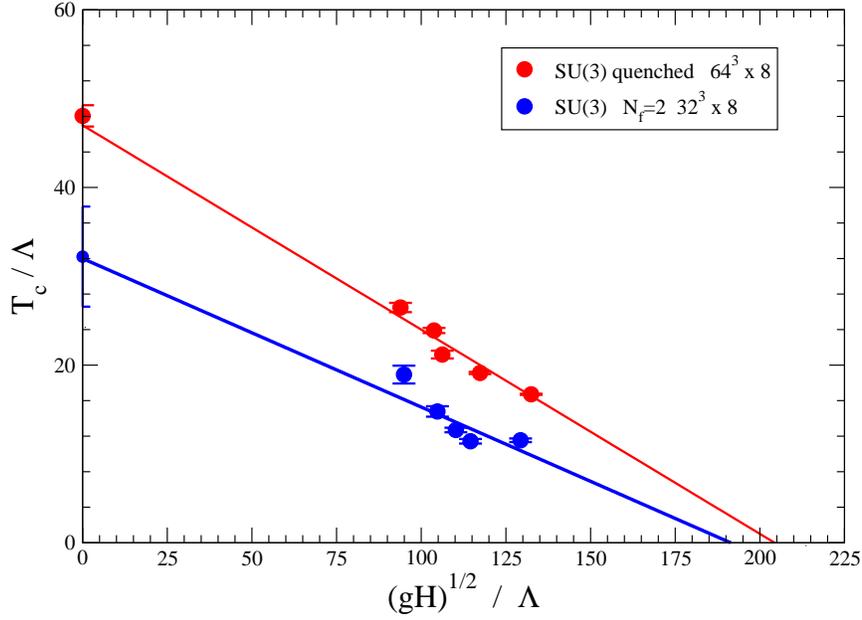
\caption{$T_c/\Lambda$ versus the strength of the applied chromomagnetic field
in unit of the scale $\Lambda$ (Eq.~\ref{Lambda}).}}

\section{The critical field strength}
\label{criticalfs}
Now we want to repeat the analysis of Ref.~\cite{Cea:2005td} in order to ascertain if
(i) the deconfinement temperature goes to zero as increasing the chromomagnetic field strength
(ii) there exists a critical field strength beyond which the system is deconfined even at
zero temperature.

The deconfinement temperature can be derived using the values of the critical coupling
obtained by locating the peak of the derivative of the free energy in correspondence of each
value of the external field strength.

In our previous study~\cite{Cea:2005td}  in pure gauge SU(3) we used the string tension as  a
physical scale (see Fig.~\ref{Fig4a}).
\be
\label{Tphys}
\frac{T_c}{\sqrt{\sigma(\beta_c)}} = \frac{1}{L_t \sqrt{\sigma(\beta_c)}} \,.
\ee
Moreover, using eq.~(\ref{a2gH}), the field strength  is
\be
\label{gHphys}
\frac{\sqrt{gH}}{\sqrt{\sigma(\beta_c)}} = \sqrt{\frac{4 \pi n_{\text{ext}}}{L_x  \sigma(\beta_c)}} \,.
\ee
The lattice data can
be reproduced by a linear fit
\be
\label{tcsqrtsigma}
\frac{T_c}{\sqrt{\sigma}} =  \alpha \frac{\sqrt{gH}}{\sqrt{\sigma}} + \frac{T_c(0)}{\sqrt{\sigma}} \,,
\ee
with $T_c(0)/\sqrt{\sigma}  = 0.643(15)$.
Where our determination for $T_c(0)/\sqrt{\sigma}$ is consistent with the determinations
$T_c/\sqrt{\sigma}=0.640(15)$  obtained
in the literature  without external field~\cite{Teper:1998kw}.
Using eq.~(\ref{tcsqrtsigma}) the critical field can now be expressed in units of the string tension
\be
\label{su3hc}
\frac{\sqrt{gH_c}}{\sqrt{\sigma}} = 2.63  \pm 0.15  \,.
\ee
Assuming $\sqrt{\sigma}=420$~MeV, eq.~(\ref{su3hc}) gives for the critical field
\be
\label{criticalfield}
\sqrt{gH_c} = (1.104 \pm 0.063) {\text{GeV}}
\ee
corresponding to $gH_c=6.26(2) \times 10^{19}$~Gauss.

In the present work we perform a
preliminary analysis of the lattice data obtained in full QCD with $N_f=2$
using  the $\Lambda$ scale introduced in~\cite{Allton:1996kr,Edwards:1998xf}
\begin{equation}
\label{Lambda}
\Lambda=\frac{1}{a} f(g^2) (1+c_2 \hat{a}(g)^2 + c_4 \hat{a}(g)^4)
\end{equation}
with
\begin{equation}
\hat{a}(g)^2 \equiv \frac{f(g^2)}{f(g^2=1)}
\end{equation}
and $f(g^2)$ the 2-loop $\beta$-function with $N_f=2$.

The data for $T_c/\Lambda$ versus the external field strength $(gH)^{1/2}/\Lambda$ are reported in Fig.~\ref{Fig4b}.
In Fig.~\ref{Fig4b} we display also the pure gauge data analysed using the same scale $\Lambda$
given in Eq.~(\ref{Lambda}).
Data reported in Fig.~(\ref{Fig4b}) shows that the deconfinement temperature goes to zero, both in
quenched and unquenched case, in correspondence of a given critical field strength.

In particular, assuming a linear dependence on  $\sqrt{gH}$
as in the quenched case
\begin{equation}
\frac{T_c(gH)}{T_c(0)} = 1 - \frac{\sqrt{gH}}{\sqrt{gH_c}}
\end{equation}
and using the value $\Lambda/\sqrt{\sigma} = 0.01338$ (Eq.~4.3) of Ref.~\cite{Edwards:1998xf}) we found that
\begin{equation}
\sqrt{gH_c}(N_f=2)  \sim \sqrt{gH_c}(\text{quenched}) \simeq 1.1 \text{GeV} \,.
\end{equation}
The value for the critical field strength in the present analysis for the
quenched case is  well consistent with the value obtained in
Ref.~\cite{Cea:2005td} using the scale of the string tension and reported here in Eq.~(\ref{criticalfield}).

\section{Conclusions}
\label{conclusions}
We studied full QCD with $N_f=2$ flavors of staggered fermions in presence of
a  constant chromomagnetic field.
We found that, as in the quenched case (see ref.~\cite{Cea:2005td}), the critical temperature
for the deconfinement phase transition depends on the strength of the applied field and
eventually goes to zero. The value of the critical field strength is $\simeq 1$~GeV.
Moreover numerical results suggest that the chiral critical temperature is consistent
with the deconfinement temperature
and both depend on the strength of the external chromomagnetic field.

Finally our numerical results shows that the value of the chiral condensate increases
with the strength of the external chromomagnetic field. This is an intriguing effect that
deserves further investigations.

%\bibliography{qcd}
\providecommand{\href}[2]{#2}\begingroup\raggedright\endgroup

\end{document}